\documentclass{elsart} 
\usepackage{graphicx} 
\usepackage{amssymb} 
\usepackage{amsmath} 
\usepackage{bm}

\begin{document} 
\begin{frontmatter} 
\title{Optimizing WIMP directional detectors} 
\author{Anne M. Green$^1$, Ben Morgan${}^2$} 
\address{${}^1$School of Physics and Astronomy, University of 
  Nottingham, University Park, Nottingham, NG7 2RD, UK.} 
\address{${}^2$Department of Physics, University of Warwick, Coventry, 
CV4 7AL, UK.} 
 
\thanks[]{anne.green@nottingham.ac.uk} 
\thanks[]{ben.morgan@warwick.ac.uk} 
 
\begin{abstract}  
 
We study the dependence of the exposure required to directly detect a WIMP 
directional recoil signal on the capabilities of a directional detector. 
Specifically we consider variations in the nuclear recoil energy 
threshold, the background rate, whether the detector measures the 
recoil momentum vector in 2 or 3 dimensions and whether or not the 
sense of the momentum vector can be determined. We find that the property 
with the biggest effect on the required exposure is the 
measurement of the momentum 
vector sense. If the detector cannot determine the recoil 
sense, the exposure required is increased by an order of magnitude for 
3-d read-out and two orders of magnitude for 2-d read-out.  
For 2-d read-out the required exposure, in 
particular if the senses can not be measured, can be significantly 
reduced by analyzing the reduced angles with the, time dependent, 
projected direction of solar motion subtracted. The background rate 
effectively places a lower limit on the WIMP cross-section to which 
the detector is sensitive; it will be very difficult to detect WIMPs 
with a signal rate more than an order of magnitude below the background rate.  
Lowering the energy threshold also reduces the required exposure, but 
only for thresholds above 20 keV.

\end{abstract} 
 
\begin{keyword} 
 
dark matter \sep directional detection 
 
 
\PACS  95.35+d\sep  
 
\end{keyword}

\end{frontmatter}

\section{Introduction} 
Weakly Interacting Massive Particle (WIMP) direct detection 
experiments aim to directly detect non-baryonic dark matter via the 
elastic scattering of WIMPs on detector nuclei~\cite{DD}, and are 
presently reaching the sensitivity required to detect the lightest 
neutralino (which in most supersymmetry models is the lightest 
supersymmetric particle and an excellent WIMP candidate). Since the 
expected event rates are very small ( ${\cal O} (10^{-5} - 1)$ counts 
${\rm kg^{-1} day^{-1}}$) distinguishing a putative WIMP signal from  
backgrounds due to, for instance, 
neutrons from cosmic-ray induced muons or natural radioactivity, is 
crucial. The Earth's motion about the Sun provides two potential WIMP 
smoking guns: i) an annual modulation~\cite{amtheory} and ii) a strong 
direction dependence~\cite{dirndep} of the event rate.  
The dependence of the differential event rate on the atomic mass of 
the detector (see e.g. Refs.~\cite{jkg,ls,kraus}) is a third possibility, 
however this would require good control of systematics for detectors 
composed of different materials. 
 
The direction dependence of the WIMP scattering rate has several 
advantages over the annual modulation. Firstly, the amplitude of the 
annual modulation is typically of order a few 
per-cent~\cite{amtheory,ls} while the event rate in the direction of 
solar motion is roughly an order of magnitude larger than that in the 
opposite direction~\cite{dirndep,ls}. Secondly, it is difficult for the 
directional signal to be mimicked by backgrounds. In most cases (a 
point source in the lab is a possible exception) a background which is 
anisotropic in the laboratory frame will be isotropic in the Galactic 
rest frame as the time dependent conversion between the lab and 
Galactic co-ordinate frames will wash out any lab specific features. 
Low pressure gas time projection chambers (TPCs), such as DRIFT ({\bf 
  D}irectional {\bf R}ecoil {\bf I}dentification {\bf F}rom {\bf 
T}racks)~\cite{drift,sean:drift} and NEWAGE~\cite{newage},  
see also Ref.~\cite{martoff}, seem to offer the best prospects for a 
detector capable of measuring the directions of sub-100 keV nuclear recoils.

Early studies found that in principle as few as 30 events would be 
required to distinguish a WIMP induced signal from isotropic 
backgrounds~\cite{copi:krauss,lehner:dir}. In reality the number of 
events, and more importantly the required exposure, will depend on the 
detector properties including the energy threshold, background event 
rate, whether the read-out measures the recoil momentum in 2 or 3 
dimensions (and if 2-d in which plane)~\cite{pap1,pap2,copi2d} and whether 
the sense (i.e. the absolute sign $+ {\vec{p}}$ or $- { \vec{p}}$) of 
the recoil momentum vector can be 
measured~\cite{pap1,pap2,copi2d}. In this paper we extend our previous 
work~\cite{pap1,pap2} by considering the effects of non-zero background and
varying energy threshold and also making a systematic comparison
of the number of events, and hence the exposure, required to
reject isotropy for benchmark
detector configurations covering the range of possibilities.
Our aim is to provide guidance for experimentalists wishing to optimize the 
detection potential of their experiment. The performance of a real 
detector will of course be more complex than our simulated detector 
and the cost (both financial and man-hours) of improvements to the 
detector performance will in reality have to be taken into account 
when formulating an optimization strategy. None the less we believe 
that our results will provide useful indications of which aspects of 
the detector performance have most effect on the ability to detect a 
WIMP directional signal.

\section{Calculating the Nuclear Recoil Spectrum} 
We use the same method for calculating the directional nuclear recoil spectrum 
as in Refs.~\cite{pap1,pap2}. In this section we explain the  
essential details, for further information see these references.

\subsection{Modeling the Milky Way WIMP halo} 
We consider the simplest possible model of the Milky Way 
(MW) halo; an isotropic sphere with density distribution $\rho \propto 
r^{-2}$. The velocity distribution of WIMPs at all 
positions within the halo is Maxwellian:  
\begin{equation}  
\label{fmax} 
f_{0}(\vec{v}) = \frac{1}{(2\pi/3)^{3/2}\sigma_{\rm v}^3} 
     \exp{\left(\frac{3|\vec{v}|^2}{2\sigma_{\rm v}^2}\right)} 
\,,  
\end{equation}  
where $\sigma_{\rm v}$, the velocity dispersion, is 
related to the asymptotic circular velocity (which we take  
to be $v_{{\rm c}} = 220 \, {\rm km \, s}^{-1}$) by $\sigma_{\rm v} = 
\sqrt{3/2}\,  v_{{\rm c}}$.  
 
This model is quite possibly not a good approximation to the real 
Milky Way, however in Refs.~\cite{pap1,pap2} we found that the number 
of events required to reject isotropy and detect a WIMP signal changed 
only weakly for observationally and theoretically motivated smooth 
halo models. It is possible (see e.g. Refs.~\cite{nonsmooth1,nonsmooth2})  
that the 
WIMP distribution on the sub-milli-pc scales which are probed by 
direct detection is composed of streams. In this case the recoil 
momentum spectrum, and hence the detectability of WIMPs, would depend 
strongly on the number and directions of 
streams passing through the solar neighbourhood.

\subsection{Modeling the Detector Response} 

In a TPC directional detector, the accuracy with which nuclear recoil
directions can be reconstructed is limited by multiple scattering of
the nucleus and diffusion of the ionisation generated along the recoil
track. To estimate this angular resolution we model a TPC filled with
0.05bar CS$_2$ gas, a 200$\mu$m pitch micropixel readout plane, a 10cm drift
length over which a uniform electric field of 1kV cm$^{-1}$ is
applied.  This gas pressure, mixture and electric field are chosen to match the
design of the DRIFT-I/II detector~\cite{sean:drift}, and the
micropixel readout is based on those described in~\cite{bellanzini}. Although
our drift length is shorter than that in DRIFT-I/II, it is chosen so
that the rms diffusion over the full 10cm drift length is
approximately equal to the pixel pitch~\cite{ohnuki}. We also caution that
micropixel readouts have not as yet been operated with CS$_2$
gas. However, our aim is to model a detector based on an advanced but
realistic high spatial resolution readout.

We use the SRIM2003~\cite{srim} package to generate sulfur recoil
tracks. The drift and diffusion of the ionisation produced along the
track is simulated using the data given in Ref.~\cite{ohnuki}, with
charge avalanches, simulated on pixels that collect drifted
ionisation. We caution that SRIM2003 was not designed to model recoils
in gaseous targets, although it predicts sulfur recoil ranges and
quenching factors (fraction of recoil energy going into ionisation) in
agreement with experimental data~\cite{drift:neutron}. However, it is
the only tool available at present for this task.

Recoil directions are reconstructed as the principal axis of the
charge distributions recorded by the pixels. In 3-d the distribution
of the difference between the primary recoil direction and the
reconstructed track direction peaks at $\sim 15^{\circ}$, decreasing
weakly with increasing energy, and has a long large angle tail. We
take this angular resolution function into account in our 3-d
Monte Carlo simulations of the recoil angle distribution. Further
details on the simulation and reconstruction process may be found in
Ref.~\cite{bm:thesis}

At present no simulations of the angular resolution function of a 
2-d detector are available. A 2-d read-out projects the recoil track 
into a plane and the effects of this combined with multiple scattering 
and diffusion will make the angular resolution a function of both the 
energy and primary recoil direction. We therefore assume perfect 
resolution for 2-d read-out and hence provide lower limits on the 
numbers of events required for the detection of a WIMP signal with a 
2-d detector.

\subsubsection{Energy threshold} 
For primary recoil energies below 20 keV the short track length (3-4 
pixels) and multiple scattering make it impossible to reconstruct the 
track direction in our simulated detector. We therefore define the energy 
threshold as the energy above which recoil directions can be 
reconstructed. It is important to note 
that this energy is higher than the energy threshold for simply 
\textit{detecting} recoils. In addition, it may be lower than the 
energy above which recoils can be \textit{discriminated} from electron 
backgrounds. These details will vary from detector to detector.

\subsubsection{Measurability of sense} 
For sub-100keV energies, the dE/dx of nuclear recoils, and hence the
ionisation density, is predicted to slowly decrease with decreasing recoil
energy. Thus the sense of a recoil track is, in principle, measurable
by determining the direction in which the ionisation density decreases
along the track. However, the ionisation density distributions
reconstructed from our simulated recoil tracks are close to uniform
due to both fluctuations in the production of ionisation and diffusion
during the drift of this ionisation to the readout plane. 
It is therefore not possible to determine the absolute signs of the simulated
reconstructed recoil vectors (i.e $+\vec{p}$ or $-\vec{p}$). 

With our earlier caveat regarding the use of SRIM2003 to model low
energy recoils, whether or not the sense of the recoils can be
measured in reality is a question that needs to be addressed with
experimental measurements. In our Monte Carlo simulations we therefore
consider both possibilities. For clarity we refer to read-outs where
the sense is measurable as vectorial, and those where it is not as axial.

\subsection{2-d read-out} 
A 2-d read-out measures the projection of the recoil momentum vector into a 
plane $P$ fixed on the Earth. Choosing an arbitrary vector $\vec{c}$ fixed in 
$P$ allows (projected) recoil directions to be described via the angle 
$\phi$ between the projected recoil and $\vec{c}$. As well as this raw 
angle, it is useful to measure the reduced angle $\phi_{red}$ between the 
projected recoil direction and the direction of solar motion projected 
into the plane:
 
\begin{equation} 
\phi_{red} = \phi - \mu_{\odot}(t) \,,
\end{equation} 
 
where $\mu_{\odot}$ is the angle between the (projected) solar motion
direction and $\vec{c}$, and $t$ is the sidereal time at which the
recoil occurred.

The degree of anisotropy in the distributions of $\phi$ and 
$\phi_{red}$, and hence the detectability of a WIMP signal, is 
dependent on the orientation of $P$, the direction $\vec{m}$ in which 
the recoil rate peaks and the Earth's spin axis $\vec{s}$. The 
anisotropy will be maximized if i) the distance between $\vec{m}$ and 
$P$, over one sidereal day, is minimized (this minimizes smearing 
caused by projection effects) and ii) the projection $\vec{m}$  
into $P$ has minimal motion relative to $\vec{c}$ (minimizing the 
smearing caused by time-averaging)~\cite{pap2,copi2d}. These 
requirements are met, for any $\vec{m}$, if the normal to $P$ is at 
$90^{\circ}$ to the spin axis of the Earth (we refer to this plane as 
the meridian plane in Ref.~\cite{pap2}). We focus on this plane for 
all analyses, but comment on the qualitative effect of changing the 
orientation.

\subsection{Background} 
Zero background is the goal of the next generation of experiments made 
from low activity materials with efficient gamma rejection and 
shielding, located deep underground ~\cite{drift2:design}. We  
investigate the effect of non-zero isotropic background by varying the ratio 
of the background and signal event rates and also, independently, the 
background rate. 
 
 \begin{figure}[]   
\includegraphics[width=13cm]{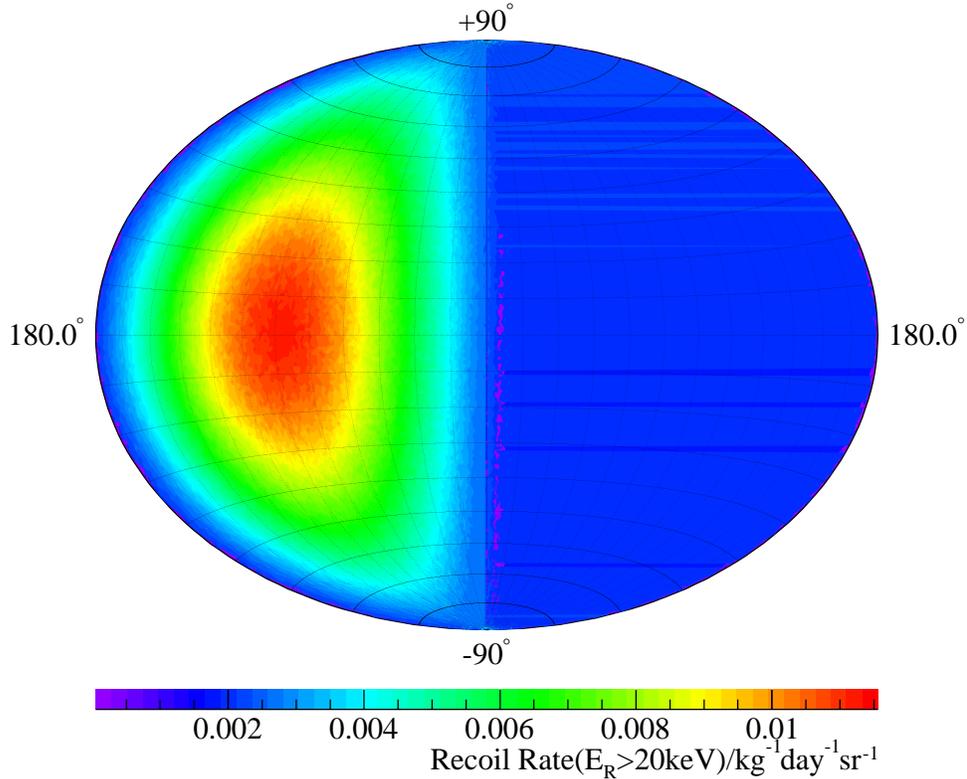} 
 \caption{The time averaged S recoil flux above 20 keV in Galactic (l,b) 
co-ordinates for the standard halo model, if the senses of the  
recoils can be measured and the uncertainty in the reconstruction 
of the recoil directions are included. The WIMP mass and cross-section 
are taken to be $m_{\chi}=100$ GeV and $\sigma=10^{-6}$ pb respectively  
and the local 
WIMP density is $\rho_{0}=0.3 \, {\rm GeV \, cm}^{-3}$. 
} 
\label{fig1} 
\end{figure}

\subsection{Summary} 
We calculate the directional recoil rate in the rest frame  
of the detector via Monte Carlo simulation of the spin-independent 
elastic scattering of $^{32}\rm S$ nuclei by WIMPs generated from 
Eq.~\ref{fmax}  with, for illustrative purposes, a mass of 
$m_{\chi}=100$ GeV. As appropriate, we include the detector angular 
resolution function and energy threshold as discussed above. For 3-d
read-out we work in coordinates of Galactic longitude $l$ and latitude
$b$. The flux distribution for sulfur recoils is plotted in Fig. 1 for a 
WIMP-nucleon elastic scattering cross-section $\sigma=10^{-6}$ pb and 
local density $\rho_{0}=0.3 \, {\rm GeV \, cm}^{-3}$.

  \begin{figure}[]   
\includegraphics[width=15cm]{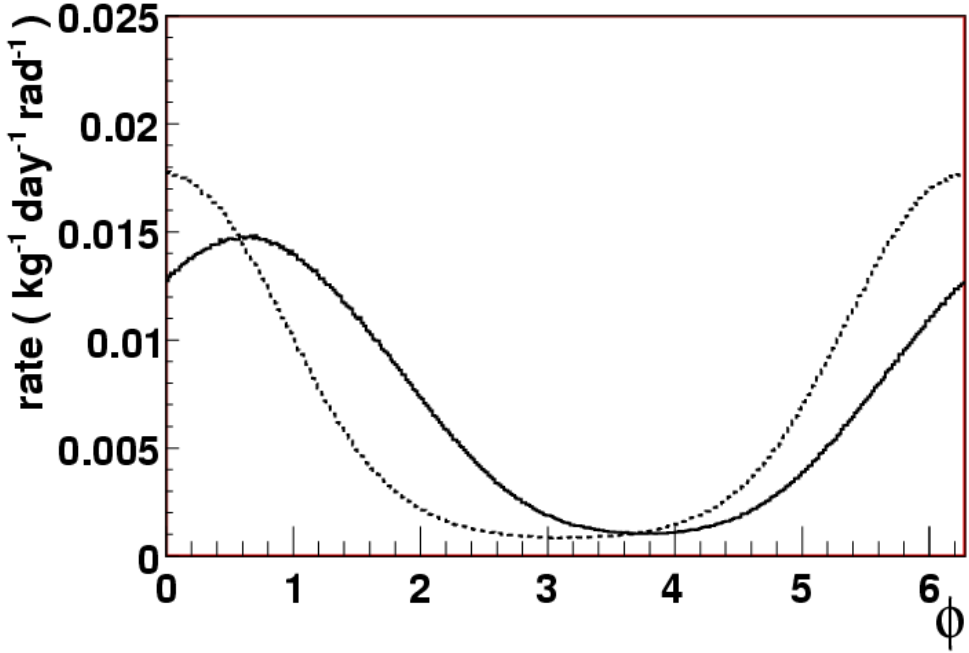} 
 \caption{The 2-d raw (solid line) and reduced (dotted line) angle 
distributions in the optimal read-out plane. The WIMP parameters are the
same as for the 3-d recoil flux distribution in Fig. 1, from which the
2-d angle distributions are generated.} 
\label{fig2} 
\end{figure}

For 2-d read-out, the distribution of the $\phi$ angles is generated
from the 3-d recoil distribution in the Galactic coordinate system by
Monte Carlo simulation of the sidereal time dependent coordinate
transformation to the detector frame, together with the 2-d projection
effects described above.  Further details of these procedures may be
found in Refs~\cite{pap1,pap2}. The raw and reduced 2-d angle
distributions are plotted in Fig. 2.

\section{Statistical Tests of Isotropy in 2-d and 3-d} 
Recoil directions in 3-d and 2-d constitute points on the unit sphere 
and circle respectively. As we showed in Refs~\cite{pap1,pap2}, 
experiments observing an anomalous recoil signal can confirm its 
Galactic nature by applying a range of simple non-parametric tests for 
anisotropy to the distribution of the observed recoil directions.

For 3-d data the most powerful test for rejecting 
isotropy uses the fact that, for smooth halo models, the WIMP recoil 
distribution is expected to be peaked about the direction of solar 
motion $(l_{\odot},b_{\odot})$. It involves calculating the average 
of the cosine of the angle between the direction of solar motion 
and the recoil direction. For the case when the sense of the recoil 
vectors are known 
\begin{equation}  
\langle \cos \theta \rangle = \frac{\sum_{i=1}^N \cos 
\theta_i}{N} \,,  
\end{equation}  
where $N$ is the number of vectors (i.e. the number of events), and if the 
recoil sense is not known then  
\begin{equation} 
\langle |\cos \theta| \rangle = \frac{\sum_{i=1}^N |\cos \theta_i|}{N} 
\,,  
\end{equation}  
where $\theta_i$ is the 3-d angle 
between $(l_{\odot},b_{\odot})$ and the $i$th vector/axis.  For  
isotropic vectors $\langle \cos \theta \rangle \, (\langle |\cos 
\theta| \rangle$) can take values on the interval $[-1,1] \, ([0, 1])$ 
and, due to the central limit theorem, approaches a gaussian distribution 
with mean $0 \, (0.5)$ and variance $1/3N \, (1/3N)$~\cite{briggs} for
large $N$. 
The larger the concentration of recoil directions towards 
$(l_{\odot},b_{\odot})$ the larger these statistics will be. 
The probability distribution function of $ \langle 
\cos \theta \rangle$ and $\langle |\cos \theta| \rangle$ as a function 
of $N$ for the null hypothesis of an isotropic recoil spectrum has 
to be calculated by Monte Carlo simulation for small values of $N$. 
 
For 2-d data we found in Ref.~\cite{pap2} that the most 
powerful test is the Rayleigh test which uses the mean resultant 
length of the projected recoil vectors which, modulo fluctuations, should 
be zero for data drawn from an isotropic distribution.   
For 2-d vectors, parameterized in terms of the 
angle $\phi$ (relative to some arbitrary fiducial direction), if we define  
\begin{eqnarray}  
C &=& \sum_{i=1}^{N} \cos{\phi_{i}} \,, \\  
S &=& \sum_{i=1}^{N} \sin{\phi_{i}} \,,  
\end{eqnarray}  
then the resultant length of the sum of the vectors, ${\cal R}$, is 
given by ${\cal R}^{2}= C^2 + S^2$ and the mean resultant direction is  
$\bar{{\cal R}}={\cal R}/N$. The 
modified 
Rayleigh~\cite{rayleigh} statistic ${\cal R}^{\star}$,  
defined as~\cite{cf,mardia:jupp} 
\begin{equation} 
{\cal R}^{\star} = \left( 1 - \frac{1}{2N} \right) 2 N \bar{{\cal R}}^2 + 
        \frac{N \bar{{\cal R}}^4}{2} \,. 
\end{equation} 
has the advantage of approaching 
its large $N$ asymptotic distribution for smaller values of $N$ than  
$\bar{{\cal R}}$. 
Under the null hypothesis that the distribution 
from which the sample of angles is drawn is isotropic, ${\cal R}^{\star}$ 
is asymptotically distributed as $\chi_{2}^{2}$  
with error of order $N^{-1}$~\cite{mardia:jupp}.  
With axial data (i.e. unsigned lines) the standard 
procedure~\cite{axes,fisher} is to double the axial angles, reduce 
them modulo $360^{\circ}$ and analyze the resulting vector data.

For each detector configuration we calculate the probability 
distribution of the relevant statistic, for a given number of events 
$N$, by Monte Carlo generating $10^5$ experiments each observing $N$ 
recoils drawn from our simulated 3-d or 2-d distributions.   
We then compare this with the null distribution of the 
statistic, under the assumption of isotropy calculated using the 
analytic expression for the 2-d Rayleigh test and by Monte Carlo 
simulation for the 3-d $\cos{\theta}$ test. Specifically we calculate 
the rejection and acceptance factors, $R$ and $A$, at each value $T$ of 
the statistic. The rejection factor is the probability of measuring a 
value of the statistic less than $T$ if the null (isotropic) 
hypothesis is true or equivalently the confidence with which the null 
hypothesis can be rejected given that measured value of the statistic. 
The acceptance is the probability of measuring a value of the 
statistic larger than $T$ if the alternative hypothesis is true or 
equivalently the fraction of experiments in which the alternative 
hypothesis is true that measure a larger absolute value of the test 
statistic and hence reject the null hypothesis at confidence level 
$R$. Clearly a high value of $R$ is required to reject the null 
hypothesis. A high $A$ is also required, otherwise any one experiment 
might not be able to reject the null hypothesis at the given $R$ or 
the null hypothesis might be erroneously rejected. We therefore find, 
using a search by bi-section, the number of events required for 
$A_{\rm c}=R_{\rm c}=0.9$ and $0.95$. A single experiment could in reality
be lucky/unlucky and need  less/more events to detect a signal, but
this procedure gives a well defined indication of the number
of events which will be required.
For further details see Appendix C of Ref.~\cite{pap1}.

\section{Results} 
 
As our baseline optimistic, but realistic, detector configuration we 
consider a detector with 3-d vector (i.e. with recoil sense
measurable) read-out, 
energy threshold $E_{\rm TH}= 20$ keV, no background ($S/N=\infty$) 
and with the uncertainty in reconstructing the recoil directions taken 
into account. One or two of these specifications are varied in each of 
the alternative configurations. We 
consider two improved (and unrealistic) configurations, one with 
$E_{\rm TH}= 0$ keV and one with perfect recoil reconstruction (since 
the recoil 
reconstruction can not be taken into account for 2-d read-out, the 
number of events required for 2-d read-out should be compared with this 
later 3-d configuration).  We also consider a number of `degraded' 
configurations: 2-d read-out, axial read-out, a larger energy threshold 
and non-zero background. For the 2-d read-out we 
consider both the raw angles and the reduced angles (with the projected  
direction of solar motion subtracted). We parameterize non-zero background  
in two different ways: variable signal to noise and variable background rate. 
Throughout we assume, for illustrative purposes, a WIMP mass of $m_{\chi} = 
100  \, {\rm GeV}$.

\begin{table*} 
\begin{center} 
\begin{tabular}{|l|c|c|c|} 
\hline 
 {\rm difference from baseline configuration} & $N_{90}$ 
 & $N_{95}$\\ 
\hline 
\hline 
  {\rm none}   & 7 & 11 \\ 
\hline 
 $E_{\rm T}= 0$ \, keV & 13 & 21 \\ 
  no recoil reconstruction uncertainty & 5 & 9 \\ 
\hline 
 $E_{\rm T}= 50$ \, keV & 5 & 7 \\ 
 $E_{\rm T}= 100$ \, keV & 3& 5 \\ 
  $S/N=10$ & 8 & 14 \\  
  $S/N=1$ & 17 & 27  \\ 
 $S/N=0.1$ & 99 & 170  \\ 
  3-d axial read-out & 81 &130 \\ 
  2-d vector read-out in optimal plane, raw angles & 18 & 26 \\ 
  2-d axial read-out in optimal plane, raw angles & 1100 & 1600 \\ 
  2-d vector read-out in optimal plane, reduced angles & 12  & 18 \\ 
  2-d axial read-out in optimal plane, reduced angles & 190 & 270 \\ 
\hline 
\end{tabular} 
\end{center} 
\caption{The dependence of the  
number of events {\em above the energy threshold} 
required to reject isotropy for $A_{\rm 
c}=R_{\rm c}=0.9$ and $0.95$, $N_{90}$ and $N_{95}$, on the  
detector configuration. The baseline configuration has 3-d vector read-out,  
energy  threshold $E_{\rm 
TH}= 20$ keV, no background ($S/N=\infty$) and the uncertainty in 
reconstructing the recoil directions taken into account.   
The second and third lines are (unrealistic) improved configurations.  
The subsequent lines are degraded configurations. 
For 
the non-zero background configurations the numbers given are the number 
of {\em signal} events required. Note that the exposure required to reject
isotropy depends on the event rate which decreases with increasing
energy threshold (see figs. 3 and 4 and main text for further discussion).
} 
\end{table*}

Table 1 gives the details of the configurations considered and the 
number of events {\em above the energy threshold} required 
to reject isotropy (and hence detect a WIMP 
signal) at 90$\%$ (95$\%$) confidence in 90$\%$ (95$\%$) of 
experiments. The variable background rate configurations are not 
included in this table as in this case the number of events depends 
on the signal rate, and hence the cross-section. 
For the non-zero background cases we use a slightly 
different procedure to that described above,  
since the background and signal are both Poisson 
processes.  In these cases for each value of the cross-section, we 
search for the required exposure by using the exposure, cross-section 
and signal to noise ratio (or background rate) to 
calculate the mean number of signal and 
background events. We then draw the number of signal and isotropic
background 
events for each of the $10^{5}$ experiments from Poisson distributions 
and calculate both the null and alternative distributions of the test 
statistic by Monte Carlo simulation. This procedure is not rigorously 
justified from a statistical point of view, but should not introduce 
significant bias/error.  
 
\begin{figure}[t!]   
\includegraphics[width=13cm]{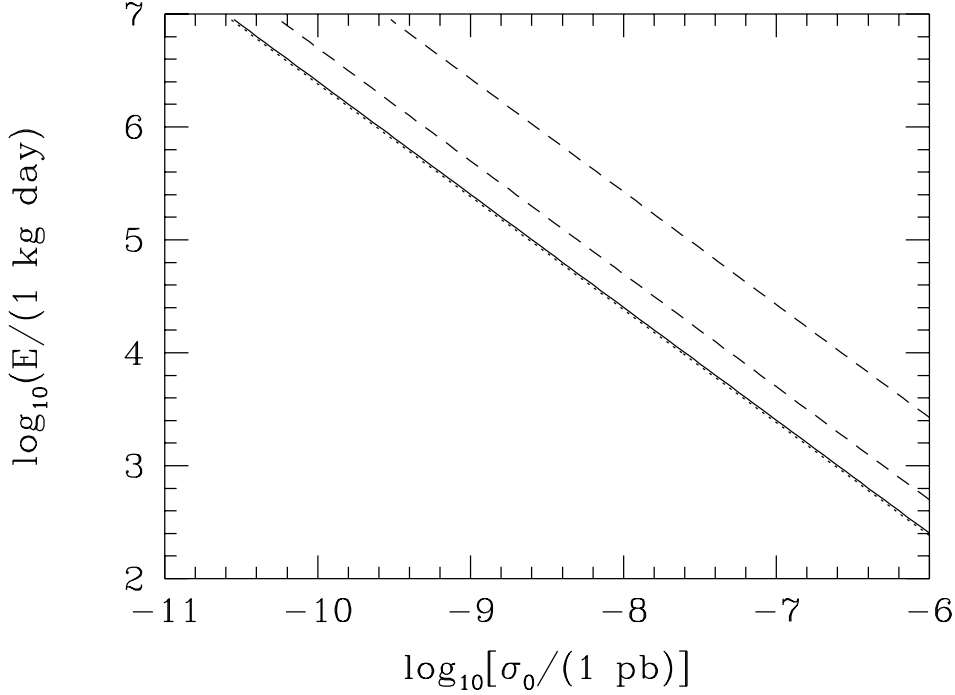} 
 \caption{The exposure required to reject isotropy (and detect a WIMP  
signal) at $95\%$ confidence in $95\%$ of experiments as a function of  
the WIMP-proton elastic scattering cross-section 
$\sigma_{0}$, assuming local WIMP density  $\rho= 
 0.3 \, {\rm GeV \, cm}^{-3}$ and WIMP mass $m_{\chi}=100 \, {\rm GeV}$.
 The solid line is for  
the benchmark configuration (zero background, $E_{\rm TH}= 20 \, {\rm keV}  
$, 3-d vector read out). The dotted lines is $E_{\rm TH}= 0 \, {\rm keV}$ 
and the dashed lines are, from bottom to top, 
$50$ and $100 \, {\rm keV}$.  
} 
\label{expos1} 
\end{figure}  

In Figs. 3-7 we plot the exposure required to accumulate the required 
number of events for a 95$\%$ confidence detection in 95$\%$ of 
experiments as a function of of the WIMP-nucleon elastic scattering 
cross-section $\sigma_{0}$ using the relationship 
 
\begin{equation} 
E= \frac{N}{\sigma_{0} R(>E_{\rm TH})} \,, 
\end{equation} 
 
where $R(>E_{\rm TH})$ is the total event rate above threshold energy, 
per unit cross-section, assuming a fiducial 
local WIMP density  $\rho_{0}= 0.3 \, {\rm GeV \, cm}^{-3}$ (the required 
exposure is inversely proportional to the local density). 
For the benchmark value $E_{\rm TH}= 20 \, {\rm keV}$,  
$R(>E_{\rm TH})= 4.5 \times 10^{4} \, {\rm kg}^{-1} \, 
{\rm day}^{-1} \, {\rm pb}^{-1} $, while 
 for $E_{\rm TH}= 0, 50$ and $100 \, {\rm keV}$, 
$R(>E_{\rm TH}) = 8.7 \times 10^{4}, 1.4 \times 10^{4}$ and $ 
1.9 \times 10^{3} \, {\rm kg}^{-1} \, 
{\rm day}^{-1} \, {\rm pb}^{-1}$ 
 respectively~\footnote{The 
event rates, especially for large threshold energies, depend on the Galactic 
escape velocity which is not well known. This dependence is not particularly 
strong and does not effect the conclusions of this work significantly.}. 
In each figure the solid line is the benchmark configuration 
(zero background, $E_{\rm TH}= 20 \, {\rm keV}  
$, 3-d vector read out), dashed lines are degraded configurations 
and dotted lines (where relevant) are unrealistic, optimistic configurations. 
 
 \begin{figure}[t!]   
\includegraphics[width=13cm]{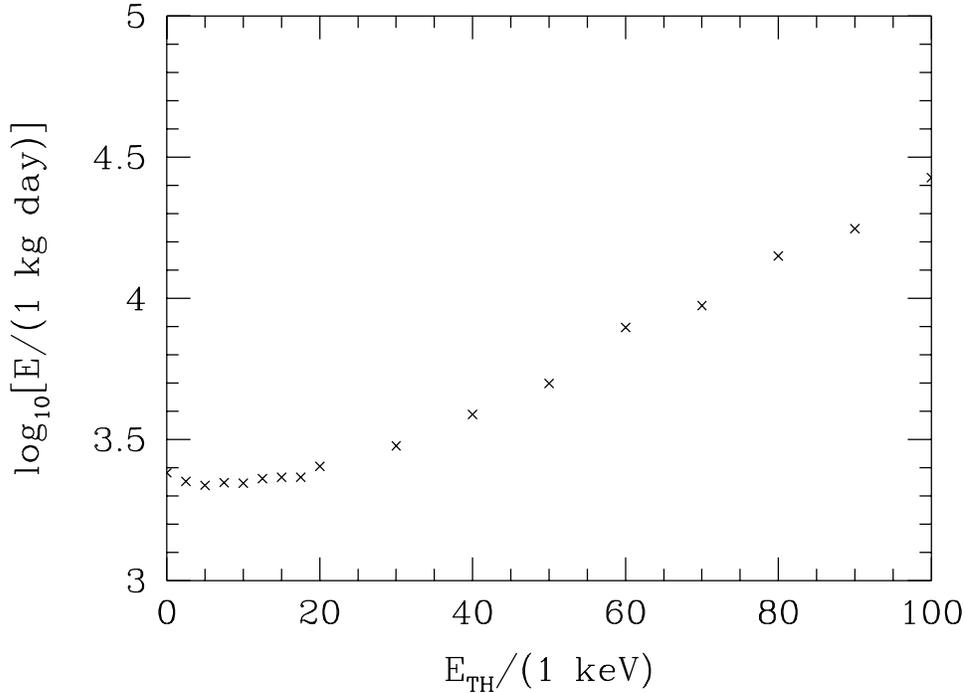} 
 \caption{The exposure required to reject isotropy (and detect a WIMP  
signal) at $95\%$ confidence in $95\%$ of experiments as a function of  
energy threshold, for  WIMP-proton elastic scattering cross-section 
$\sigma_{0}=10^{-7} {\rm pb}$, assuming a local WIMP density of $\rho= 
 0.3 \, {\rm GeV \, cm}^{-3}$. 
} 
\label{expos1b} 
\end{figure} 
 
Figs.~3 and 4 show the effects of varying the energy threshold on the 
exposure required to reject isotropy. Fig. 3 shows the variation of 
the exposure with cross-section for four values of the energy 
threshold ($E_{\rm TH}=0, 20, 50$ and $100 \, {\rm keV}$) while Fig. 4 
shows the variation with threshold energy for cross-section 
$\sigma_{0}=10^{-7} {\rm pb}$. As was pointed out by Spergel in the 
first paper on WIMP directional detection~\cite{dirndep} there are two 
competing effects. The anisotropy of the recoils increases with 
increasing energy, so that the number of events required decreases, 
however the recoil rate also decreases. We find that the net effect is that 
the required exposure is fairly constant for $E_{\rm TH} < 20 \, {\rm 
keV}$ and then increases with increasing threshold energy. As the number of 
events required is an integer the detailed variation of the exposure,
in particular at large threshold energies where there are step-like
variations, is not particularly significant. As 
a rough indication, above $20 \, {\rm keV}$ a $10 \, {\rm keV}$ 
decrease in the energy threshold decreases the exposure required by 
roughly $25\%$. The quantitative variation of the exposure will depend 
on the detector composition, however we expect that this qualitative 
behaviour is fairly generic.  Note that we have assumed in Fig. 4 that 
the recoil directions can be reconstructed down to $E = 0 \, {\rm 
keV}$. In reality it will be difficult, if not impossible, to 
reconstruct the direction of low energy recoils due to their short 
path-lengths. 
 
\begin{figure}[t!]   
\includegraphics[width=13cm]{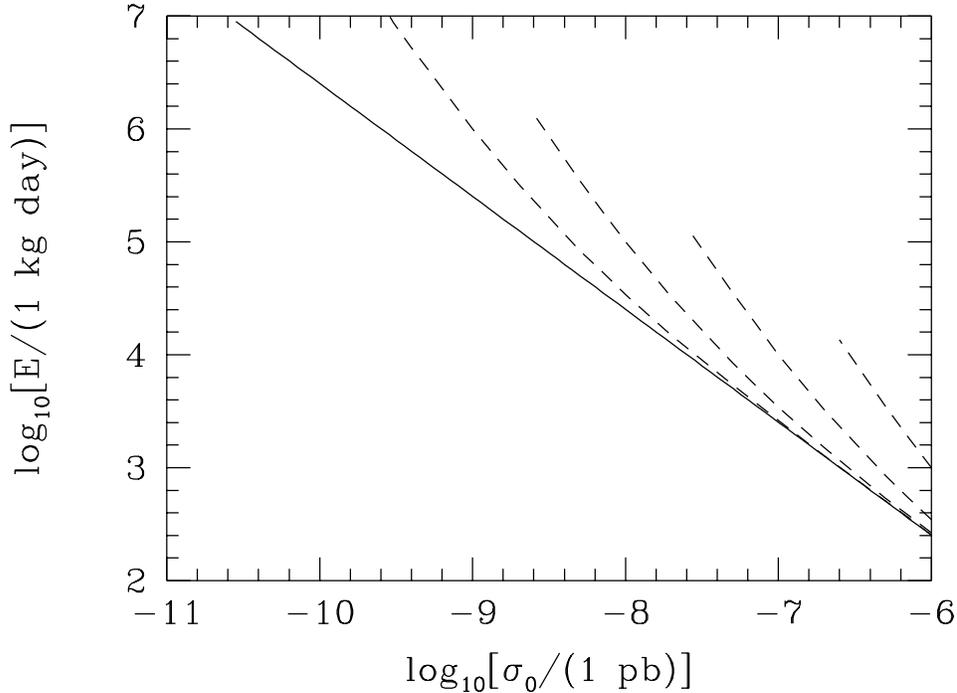} 
 \caption{As fig. 3.  The solid line is for  
the benchmark configuration (zero background).  
The dashed lines are, (from right to left), 
background event rates of $=0.1, 0.01, 0.001$ and $0.0001 \, {\rm kg}^{-1} 
{\rm day}^{-1}$.} 
\label{expos2} 
\end{figure} 

Fig. 5 shows the variation of the exposure with cross-section for 
background rates of $0, 0.1, 0.01, 0.001$ and $0.0001 {\rm kg}^{-1} {\rm 
day}^{-1}$. In each case the exposure curve follows the 
zero-background exposure curve at large cross-sections (where the 
signal rate is much larger then the background rate) before rising 
above the zero background curve, at first (when the signal and 
background rates are comparable) gently and then (once the background 
rate is much greater than the signal rate) more rapidly. For fixed 
signal to noise ratio the number of events required is independent of 
cross-section and the corresponding exposure lines would lie parallel 
to the zero-background exposure line. When the background 
rate is the same as the signal rate 
the number of events, and hence the exposure, required 
is increased by a factor of  $\sim 2.5$. 
When the background is an order of magnitude bigger than the signal rate 
the increase is a factor of  $\sim 15$. 
 
 \begin{figure}[t!]   
\includegraphics[width=13cm]{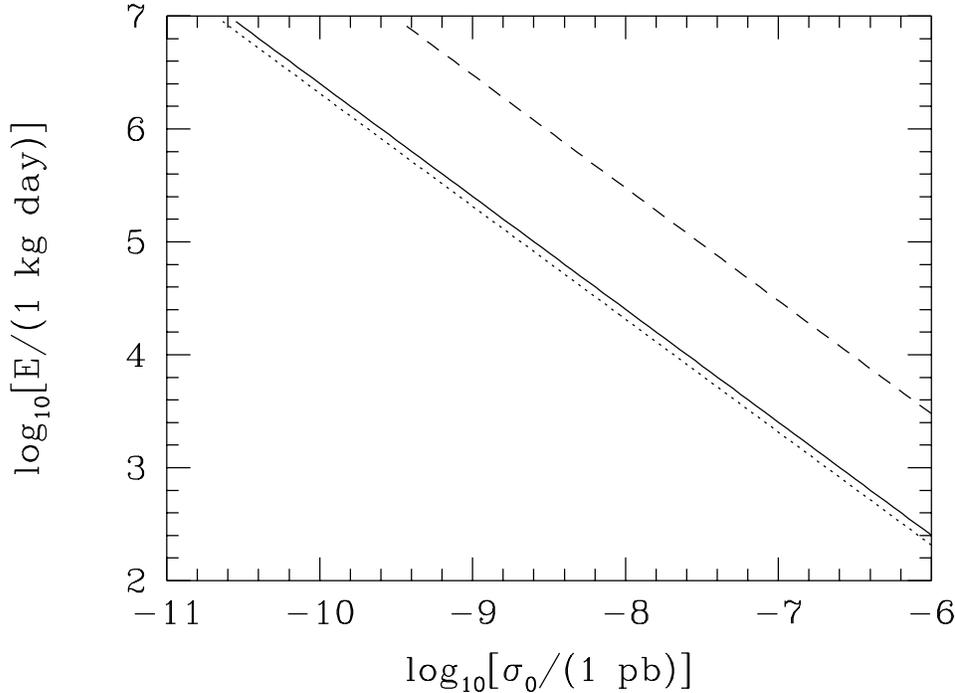} 
 \caption{As fig. 3.  The solid line is for  
the benchmark configuration (3-d vector read out, recoil 
reconstruction uncertainty taken into account).  
The dotted line is 3-d vector read out ignoring 
the uncertainty in the reconstruction of the recoil direction. 
The dashed line is axial 3-d read-out.} 
\label{expos3} 
\end{figure} 
 
Fig. 6 shows the effects of the recoil reconstruction uncertainty and 
axial read-out in 3-d. The improvement which would come from perfect 
recoil reconstruction is relatively small, of order ten per-cent. If 
the read-out is axial rather than vectorial however, the exposure 
required is increased substantially, by more than an order of 
magnitude.

\begin{figure}[t!]   
\includegraphics[width=13cm]{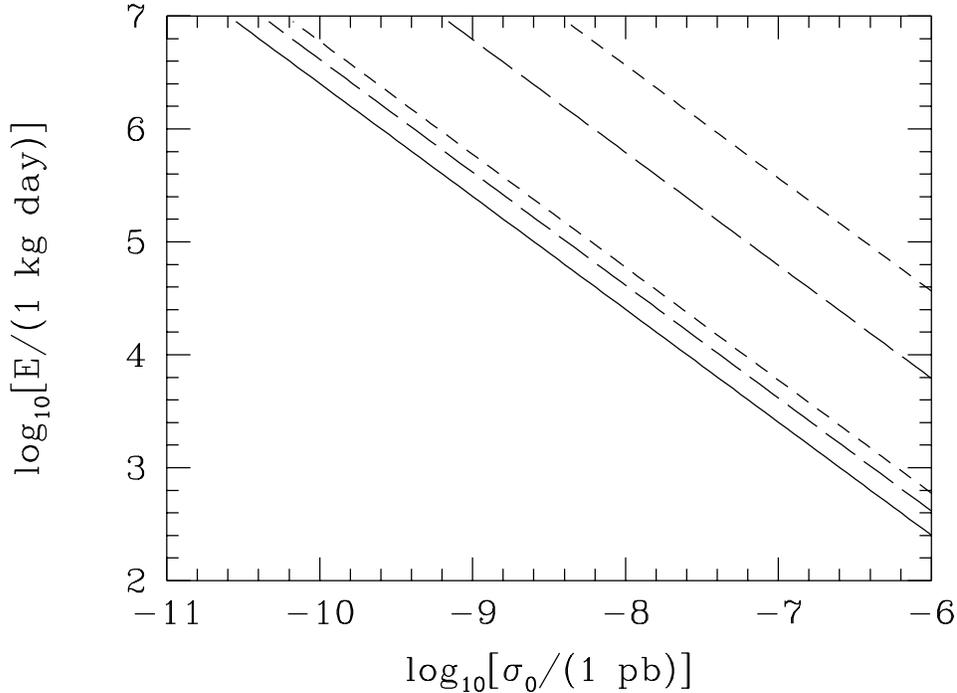} 
 \caption{As fig. 3.  The solid line is for  
the benchmark configuration (3-d vector read-out, recoil 
reconstruction uncertainty taken into account).   
The short/long dashed lines are using raw/reduced angles for 
2-d axial   
(top pair of lines) and vector  (lower pair of lines) read-out. 
} 
\label{expos4} 
\end{figure}

Fig.~7 compares 2-d read-out (vector and axial, and using raw and 
reduced angles) with 3-d read out. When the raw angles are used the 
exposure required is increased by a factor of $\sim 3$ for vectorial 
data, and by two orders of magnitude for axial data.  Smaller 
exposures, in particular for axial data, can be achieved by using the 
reduced angles, which are measured relative to the (projected) 
direction of solar motion. Relative to 3-d vector read-out, the 
exposure using reduced angles is increased by a factor $\sim 2/30$ for 
2-d vector/axial read-out. These numbers are for read-out in the 
optimum plane, and are independent of the detector's geographical 
location. Read-out in other planes would lead to larger exposures by 
up to a factor of a few for vectorial raw and reduced angles and axial 
reduced angles and up to an order of magnitude for axial raw angles 
~\cite{pap2}. As the geographic location of the detector influences 
the orientation of these planes, the exposures for non-optimal planes
vary by of order tens of per-cent between Boulby, Sudbury and
Kamioka~\cite{pap2}. We should caution that the results for 2-d
read-out assume a detector with perfect angular resolution. The
projective effects of a 2-d read-out will make this a more significant
factor than for 3-d read-out.

\section{Conclusions} 
 
We have examined the effects of the detector properties on the number 
of events, and hence exposure, required to reject isotropy of recoil 
directions, and hence identify a putative WIMP signal as Galactic in 
origin. Specifically we have 
considered the nuclear recoil energy threshold, the background rate, 2-d and 
3-d read-out and whether or not the sense of the recoils can be 
measured.  Our simulations are for a TPC detector filled with 
$CS_2$. We expect that results for other directional detectors would 
be qualitatively similar.

We took a detector with 3-d vector read-out, energy threshold $E_{\rm 
TH}= 20$ keV, no background ($S/N=\infty$), with the uncertainty in 
reconstructing the recoil directions taken into account, as our 
baseline realistic detector configuration and then varied one or two 
of these properties at a time. The property which has the largest 
effect on the exposure required is whether or not the sense of the 
recoils can be measured. If the data is axial (i.e. the sense can not 
be measured) the exposure is increased by one order of magnitude for 
3-d read-out and two orders of magnitude for 2-d read-out compared to 
the baseline detector. With 2-d read-out this can be improved to a 
factor $\sim 30$ if the reduced angles (with the direction of solar 
motion subtracted) are calculated and analyzed. If the read-out is 
2-d, and the senses can be measured, the exposure is increased by a 
factor of $\sim 3$ over the baseline. Once more this can be reduced, 
to a factor of $\sim 2$, by using the reduced angles. These figures 
are for 2-d read-out in the optimal plane, which has normal perpendicular 
to the Earth's spin axis. For other read-out planes the required 
exposures are larger, and location dependent (see section 3 and 
Ref.~\cite{pap2} for further information). We also caution that the
results for 2-d read-out assume perfect angular resolution, and projection
effects will make this a more significant factor than for 3-d
read-out. Estimates based on the projected length of recoil tracks in
the planes indicate that the required exposure would increase by at least a
factor 2~\cite{pap2}.

The effect of non-zero background depends on the underlying signal 
rate.  If the signal and background rates are the same the exposure, 
relative to that for zero background, is increased by a factor of 
$\sim 2.5$. If the background rate is an order of magnitude larger 
than the signal rate the increase is a factor of $\sim 15$. As the 
signal to noise ratio is reduced below 0.1 the detection of the 
signal anisotropy becomes extremely challenging. 
 
As the energy threshold is increased the WIMP recoil signal becomes 
increasingly anisotropic and the number of events required 
decreases. However the event rate decreases and the net effect is that 
the exposure required increases with increasing energy threshold above 
$E_{\rm TH} \approx 20 \, {\rm keV}$. The exposure increases by a factor 
$\sim 1.3$ for every $10 \, {\rm keV}$ increase in the energy threshold. 
Below $E_{\rm TH} \approx 20 \, {\rm keV}$ the exposure is roughly 
independent of the threshold energy. In practice achieving such low 
energy thresholds is not possible; the path lengths of low energy 
recoils are too short for the directions to be measured. We also 
caution that the actual `energy threshold' of a directional detector 
is the maximum of the detection, discrimination and direction 
reconstruction thresholds.

In summary the property with the biggest effect on the detector 
performance is whether or not the sense of the recoils can be 
measured. For 2-d read-out the performance, in particular if the 
senses can not be measured, can be significantly improved by recording 
and analyzing the reduced angles (with the projected direction of solar motion 
subtracted). The background rate effectively places a lower limit on 
the WIMP cross-section to which the detector is sensitive; it will be 
very difficult to detect a WIMP directional signal with a recoil rate 
more than an order of magnitude below the background rate. If the 
energy threshold is larger than about 20 {\rm keV} significant 
improvements in the exposure required can be made by reducing the 
energy threshold (i.e. a factor of 2 if $E_{\rm th}$ is reduced from 
$50 \, {\rm keV}$ to $20 \, {\rm keV}$ and a factor of 5 if it is 
reduced from $100 \, {\rm keV}$ to $50 \, {\rm keV}$).

\end{document}